# First results from a Dark Matter search with liquid Argon at 87 K in the Gran Sasso Underground Laboratory.

P. Benetti[a], R. Acciarri[f], F. Adamo[b], B. Baibussinov[g], M. Baldo-Ceolin[g], M. Belluco[a], F. Calaprice[d], E. Calligarich[a], M. Cambiaghi[a], F. Carbonara[b], F. Cavanna[f], S. Centro[g], A.G. Cocco[b], F. Di Pompeo[f], N. Ferrari[c] [†], G. Fiorillo[b], C. Galbiati[d], V. Gallo[b], L. Grandi[a], A. Ianni[c], G. Mangano[b], G. Meng[g], C. Montanari[a], O. Palamara[c], L. Pandola[c], F. Pietropaolo[g], G.L. Raselli[a], M. Rossella[a], C. Rubbia[a](+), A. M. Szelc[e], S. Ventura[g] and C. Vignoli[a]

*(a) Dipartimento di Fisica Nucleare e Teorica, INFN and University of Pavia, Italy*
*(b) Dipartimento di Scienze Fisiche, INFN and University Federico II, Napoli, Italy*
*(c) Laboratori Nazionali del Gran Sasso dell'INFN, Assergi (AQ), Italy*
*(d) Department of Physics, Princeton University, Princeton, New Jersey, USA*
*(e) Instytut Fizyki Jadrowej PAN, Krakow, Poland*
*(f) Dipartimento di Fisica, INFN and University of L'Aquila, Italy*
*(g) Dipartimento di Fisica, INFN and University of Padova, Italy*

**(WARP Collaboration)**

**Abstract.**

A new method of searching for dark matter in the form of weakly interacting massive particles (WIMP) has been developed with the direct detection of the low energy nuclear recoils observed in a massive target (ultimately many tons) of ultra pure Liquid Argon at 87 K. A high selectivity for Argon recoils is achieved by the simultaneous observation of both the VUV scintillation luminescence and of the electron signal surviving columnar recombination, extracted through the liquid-gas boundary by an electric field.

First physics results from this method are reported, based on a small 2.3 litre test chamber filled with natural Argon and an accumulated fiducial exposure of about 100 kg x day, supporting the future validity of this method with isotopically purified $^{40}$Ar and for a much larger unit presently under construction with correspondingly increased sensitivities.

*We would like to dedicate this paper to the memory of our colleague and friend Nicola Ferrari.*

# 1.— Introduction.

Recent important results based on cosmic microwave background, supernova and gravitational lensing studies have strengthened the evidence of a non-baryonic dark matter component $\Omega_{nb}$ in the Universe [1-4]. Such a Standard Model of cosmology predicts $\Omega_{nb} \approx 0.23$, well above the matter of baryonic origin, for which the value $\Omega_b \approx 0.04$ refers to Big Bang Nucleo Synthesis [5].

A satisfying explanation for this dark matter puzzle is provided by Weakly Interacting Massive Particles (WIMP). Most super-symmetric Models (SUSY) naturally offer a suitable WIMP candidate in the form of their lightest supersymmetric particles (neutralinos), provided they have survived cosmological decay, as for instance ensured by conservation of R-parity [6-10]. These models predict a wide range of possible masses and cross sections, which may become accessible to direct detection experiments. In the direct search for such particles, one looks for nuclear recoils induced by scattering on a target of WIMP that are part of the dark matter halo of our Galaxy [11]. The recoil energies range from a few keV to a few tens of keV, a relatively low energy scale for usual particle physics. These rare events must be discriminated from the much larger background rate from natural radioactivity.

Up to now, the best sensitivities have been obtained by detectors operating at the very low temperatures of a few tens of mK. A heat (or phonon) channel measures the energy deposit independently of the nature of the recoiling particle. A second channel measures the ionisation yield in a semiconductor crystal (CDMS [12,13] and EDELWEISS [14,15]) or the light yield of a scintillating crystal (CRESST [16]). The backgrounds from $\beta$ and $\gamma$ radiation are reduced by the fact that electron recoils have larger ionisation or scintillation yields than nuclear recoils.

These very low temperature experiments, which have already achieved sensitivities well in the range of predictions of SUSY/neutralino, have given no evidence for a WIMP-like signal. Within the standard theoretical framework [11], their results are in contrast with the previous ones from the DAMA experiment [17] in the Gran Sasso Laboratory, which has claimed a strong positive evidence (>99% confidence level) of WIMP due to the periodic variation of counting rate induced by the tiny yearly speed modulation of the halo galactic motion with respect to the Earth.

The detection method based on pure noble cryogenic liquids (Xenon, Argon, Neon) allows to work at much higher temperature (165 K for Xenon, 87 K for Argon and 27 K for Neon) and to be sensitive to WIMP-induced nuclear recoils through the simultaneous measurements of both the prompt scintillation and the delayed ionisation signals. First attempts to distinguish heavy ion recoils from $\beta$ and $\gamma$ radiation in Xenon were reported in 1993 [18], in connection with the research developments of ICARUS liquid Argon TPC for underground physics. Active developments have been extended since that time. Subsequently we have opted for Liquid Argon[1] (LAr), which has shown, as we shall describe later on, much better characteristics for the signal, easier availability and lower cost.

The aim of the experiment is the one to develop a unit with more than 100 litres of active volume. We report in this paper some early physics results, based on a small test chamber and an accumulated fiducial exposure of about 100 kg x day.

This preliminary search for WIMP events has been performed at the Gran Sasso Laboratory with a small 2.3 litre table-top test chamber (Figure 1) entirely surrounded by a passive absorption shield made of lead and of polyethylene. We presently describe in detail the detection method and report a first upper limit for the WIMP search with sensitivity comparable to the best published result with temperatures of few tens of mK [12] (CDMS).

In analogy with CDMS and EDELWEISS, the 2.3 litre table-top test chamber shows a persisting, tiny residual signal of events, in our case likely ascribed either to residual neutron background from natural radioactivity or spurious events due to statistical fluctuations of the relatively high level of general background (about 6 counts/s). In the future such backgrounds will be very strongly reduced since the purification of the liquid Argon will be considerably improved and the detector will be entirely surrounded by an active anticoincidence volume of as much as 8000 kg of liquid Argon. This will allow to separate WIMP events from the background due to neutrons which should generally produce one or more iso-lethargic interactions in the anticoincidence surrounding the active detector.

Natural Argon is made of spinless isotopes. Therefore the primary elastic WIMP scattering on Argon nuclei is expected to behave as a coherent cross-section, proportional to the square of the number of nucleons. Although the recoil kinetic energy of the ion is very small, the momentum transfer is quite significant and important effects are due to the nuclear form factor.

The recoil energy spectrum from WIMP scattering has been calculated using the formula and the prescriptions of Ref. [11]. A spherical isothermal halo of WIMP with a local density of 0.3 GeV/c$^2$/cm$^3$ is assumed, with an escape velocity of 600 km/s and an Earth-halo most probable speed [11] $v_o$ = 230 km/s. The spectrum is multiplied by the form factor for coherent scattering [19].

---

[1] Very large masses, up to 600 ton of instrumented Argon, have been operated within the framework of the ICARUS programme.

Our result confirms the previous conclusion that the published results of the DAMA experiment [17] cannot be ascribed, in the standard framework, to the observation of WIMP, at least as long as an important contribution is due to the spin independent cross section [20].

## 2.— Experimental method.

### 2.1. Previous work.

The direct detection of WIMP related low energy nuclear recoils observed in a massive target (ultimately many tons) of ultra pure Liquid Argon at 87 K is achieved by the simultaneous observation of (1) the electron signal surviving columnar recombination, extracted through the liquid-gas boundary by an electric field and (2) the detailed shape of the VUV luminescence pulse. Several developments in different subjects have been the necessary prerequisites to this work. Some of them are here briefly mentioned.

*Electron extraction from liquid to gas.* The extraction of electrons both from liquid Argon and liquid Xenon to gas is extensively reported in the literature [21], following the original work of Dolgoshein et al. [22] in 1973 and subsequently and extensively studied also by us [23]. The extraction process depends on an emission coefficient, function of the temperature and of the local electric field. Classically, this is related to the work required to extract a negative charge from a dielectric material. In the case of liquid Argon and Xenon, this potential barrier is large compared to the electron temperature kT. Hence the spontaneous rate of emission is very small. However an applied, local accelerating electric field is capable of increasing the electron temperature to a sufficient level as to permit the quick extraction of the electrons. Therefore a grid must be inserted, still in the liquid phase, before the transition to gas, in order to increase the field in the last bit of liquid and in the gas region, in order to ensure that the role of the "heater" is assumed by the increased electric field. A higher field in the gas region is also useful for the subsequent multiplication of the electrons, once extracted. The agreement between our points [23] and those published in Ref. [22] is excellent. Fast extraction of electrons from Argon is substantially complete already at values of the field of 3 kV/cm. The extraction of electrons from Argon occurs much more easily than for Xenon, for which there is practically no fast component below 2 kV/cm: even at 5 kV/cm the extraction efficiency is only about 90% [21]. In order to obtain comparable values, the extraction field in Xenon must be about a factor 4 larger than the one needed for Argon.

*Dependence of luminescence on the ionisation density.* This phenomenon has been first noticed by Birks [24] as early as in 1964 for several liquid scintillators and routinely used to separate with the help of the differences in the pulse shape, for instance neutron-induced recoils from electrons. In organic liquids, they have been explained with spur-recombination processes [25]. At low ionization density, electrons and ions recombine through germinate recombination, which favours the production of the (faster) singlet state. At high ionization density, homogeneous recombination is the main process and electrons and ions recombine at random with a consequent depression of the singlet state, in which for instance two long-lived triplet states may collide in an organic solution producing an excited singlet state and a ground state [26].

As well known, the Birks effect [24] cannot be directly applied to the case of noble liquids, since here the faster component shows the opposite effect. Evidence for a strong ionization density dependence of the scintillation shape of Xenon and Argon was discovered by Kubota et al. [27] in 1978. The effect was later studied by many groups [28], which have observed dual component time distributions for electrons, alpha particle and fission fragments, with and without a drift electric field of up to 4 kV/cm. Hitachi et al. [29] have published their most complete work in 1983. It remains for us as the most relevant reference of the dependence of ionization density on the time shape of luminescence of liquid Argon and Xenon.

The luminescence signals both from Xenon and Argon have been extensively studied over many years by the ICARUS collaboration [30]. It was concluded that some of the differences observed by the early measurements [28] were likely induced by electro-negative impurities, well under control both in the work of Hitachi et al. [29] and in our measurements. In particular the slow time component in Liquid Argon has been accurately measured [31]. It was also observed that even a very small amount (<< 0.01 %) of Xenon added to Argon was sufficient to shift completely the scintillation light both in wavelength and in time shape from Argon to Xenon [32], implying the presence of a full collisional transfer from excited Argon to excited Xenon. The transparency of Argon to its own VUV light was extensively investigated by the ICARUS collaboration and found to be in excess of 0.8 m [32].

These studies were performed with visible light photo-multipliers [33] cooled down (for Argon as low as $\approx$ 90 K), after having converted the light from VUV to visible with the help of Tetra-Phenyl-Butadiene (TPB) wave-shifter coating on the walls and a fast fluorescence time of few ns. The spectral shape of the photo-multipliers signals was digitally processed with a fast ($\geq$ 20 Mc/s) flash ADC followed by a shift register memory [30] in order to store for off-line analysis with all details and without biases the complete time pattern (digital sweep) of each triggered event.

The advantages of using the so called Birks/Kubota effect to recognize heavy ion recoils from electrons both in liquid Xenon [34] and in liquid Argon [35] have also been recently described by several other authors.

Our very early studies (1992) were performed with Xenon. It was however realized already sometime ago (1999) that Argon was offering in our view a far better overall performance [36]. A main background with Argon is the presence of radioactive isotopes $^{39}$Ar and $^{42}$Ar, which can be strongly reduced with isotopic purification [39] or other equivalent methods [40].

*2.2. Experimental setup.*

The detector (Figure 1) is a two-phase drift chamber, with a lower liquid Argon volume and an upper region with Argon in the gaseous phase, both viewed by the same set of photo-multipliers (PMTs). Ionisation electrons generated in the liquid are drifted by the means of an electric field to the liquid-gas interface, where they are extracted through the boundary and detected by the proportional scintillation light generated by the electrons accelerated in a high electric field. Signals from hypothetical WIMP events are recorded by the same array of photo-multipliers that observe both the scintillation light in the liquid Argon, presenting very different time behaviours for electron-like and ion recoil events both for the prompt scintillation and for the delayed ionisation signals.

The drift volume, 7.5 cm long, is delimited by a 20 cm diameter stainless steel cathode and by a system of field-shaping electrodes that generate very uniform electric drift fields (1 kV/cm). A grid (g1) placed just below the liquid level closes the uniform drift field region, while two additional grids (g2 and g3 from bottom to top) are placed in the gas phase. The sensitive volume (1.87 litres) is further delimited by a conically shaped spacer made of PTFE and placed inside the field-shaping rings; the function of the spacer is to avoid captures of ionisation electrons by the shaping rings nearby. The electric field between g1 and g2 provides both prompt extraction of the drifting electrons from the liquid to the gas and their acceleration in the gas phase to generate a secondary proportional scintillation light. The field between g2 and g3 is set in such a way to ensure complete collection of the electrons on the last grid. The grid g3 is placed 40 mm below the PMTs' windows; the field in this region is also set in such a way to ensure the main collection of electrons on g3. The liquid-gas interface has been positioned between g1 and g2 at (5.0 ± 0.5) mm above g1. Potentials of each of the grids are set with independent supplies.

In normal data taking conditions, the chamber is operated with a drift field of 1 kV/cm, the extraction/multiplication field (field between g1 and g2, 2.75 cm spacing) is 4.4 kV/cm while the field between g2 and g3 (2.5 cm spacing) is 1 kV/cm.

Seven 12-stage 2" PMT's, manufactured in order to operate at LAr temperature and placed at about 4 cm above the last grid, detect both the primary scintillation and the proportional scintillation light. Sensitivity to VUV photons emitted by the scintillating Argon is achieved by coating the photo-multiplier window with an appropriate compound, i.e. Tetra-Phenyl-Butadiene (TPB), which acts as a fluorescent wavelength shifter of the VUV scintillation light to the photo-multiplier sensitive spectrum. Average PMT's quantum efficiency in the blue region (corresponding to the emission peak from TPB) is ≈ 18%.

In order to improve the light collection efficiency from the drift volume, a high performance diffusive reflector layer surrounds the inner drift volume and the gas volume between the top grid and the photo-multipliers (Figure 1). The reflector is glued on a Mylar™ sheet with TPB deposited on its active side. The reflectivity in the TPB emission spectral region was measured to be about 95%. The reflector and the photo-multiplier windows cover about 95% of the surface surrounding the active volume.

The system is contained in a stainless steel, vacuum-tight cylindrical vessel, 25 cm in diameter and 60 cm in height. The whole container is cooled down to about 86.5 K by an external liquid Argon bath. This set-up ensures in the inner container a constant absolute pressure few mbar above the external atmospheric pressure. Residual concentration of electronegative impurities of the Argon is of the order of ≤ 1 ppb ($O_2$ equiv.), corresponding to a free electron mean free path well in excess of 0.5 metres. An Argon recirculation system is implemented working in closed loop and providing a continuous re-purification of the Argon contained in the chamber, in order to run for long periods without appreciable variations of the free electrons lifetime. Two liquid Argon level meters allow the liquid level to be positioned in between the two lowest grids with a precision of about 0.5 mm.

The integrated signal from the anode of each photo-multiplier is split by a passive divider in two copies with relative amplitudes 1:10 which are sent to two 10 bit flash ADCs with 20 MHz sampling frequency. Memory buffers are recorded at each trigger, the event structure and data are sent to the host computer and recorded to disk.

In order to generate the trigger, signals are also derived from the 12$^{th}$ dynode of each PMT. They are amplified and discriminated with a threshold of 1.5 photoelectrons. The trigger condition is ≥ 3 PMT's with at least 1.5 photoelectrons each, corresponding to about 3.5 keV ion recoil energy. The resulting trigger efficiency is close to unity above about 20 photoelectrons (16 keV ion recoil energy).

The detector assembly is completely surrounded by a shield made of 10 cm thick Lead walls encapsulated in stainless steel boxes. Internal dimensions are 110 x 110 x 270 cm$^3$. Outside the lead shield, a polyethylene shield has been installed with an equivalent thickness of 60 cm in order to ensure an additional attenuation both for $\gamma$'s and fast neutrons coming from outside.

# 3.— Data taking.

## *3.1. Initial filling.*

The detector has been filled with good grade commercial Argon, without any "ad hoc" additional purification. The resulting trigger rate is of the order of 6 counts/s, due to several radio-nuclides that are present either in the liquid or in the surrounding materials.

In this preliminary phase of development of our novel technology, the presence of such impurities is indeed beneficial, since it permits to develop the very high rejection capability against spurious events due to backgrounds from $\gamma$ and $\beta$ radiation in a relatively short data taking time.

The energy region 2 ÷ 3 MeV is presently dominated by interactions of $\gamma$-rays from $^{232}$Th daughters, the region 1.5 ÷ 2 MeV by $\gamma$-rays from $^{238}$U daughters and the region 0.5 ÷ 1.5 MeV by $\gamma$-rays from $^{60}$Co and $^{40}$K. Below 0.5 MeV the main contribution comes from $\beta$ decays from internal contaminations of $^{39}$Ar and $^{85}$Kr, both pure $\beta$ emitters with end points at 565 keV and 687 keV respectively. The specific activity of $^{39}$Ar is 1.41 ± 0.11 Bq/litre of natural Argon. For further details we refer to Ref. [37]. The result is in agreement with the atmospheric determination reported in Ref. [38].

During the earlier part of the run, immediately after filling, a significant signal due to the presence of Radon inside the chamber has been observed, with a decay half-life of 3.8 days. The presence of such a Radon induced signal is very useful, since it has permitted a real time calibration of the scintillation light yield of heavy ion recoils.

In a future phase of the 2.3 litre test chamber, these backgrounds should be strongly reduced using liquid Argon of very high purity depleted of the radioactive $^{39}$Ar isotope, naturally present in ordinary atmospheric Argon [37, 38]. Two methods are being actively pursued to this effect. In a first approach, isotopic $^{40}$Ar separation is performed with centrifugal methods [39]. In a second, parallel approach, Argon, free of radioactive impurities, will be extracted from a deep underground stacked gas reservoir [40]. Although significant amounts of geologic Argon are current available, the content of $^{39}$Ar may still be significant, since it may be also produced as a daughter of $^{39}$K (n,p) initiated by the Th or the U radioactive chains. Therefore further investigations on the origins of geologic Argon are necessary.

## *3.2. Signal recording.*

We select events consisting of both a primary light pulse (S1), from the collection of the direct scintillation light, and one or more secondary light pulses (S2) from the proportional light produced by the ionisation electrons after extraction from the liquid into the gas phase. Secondary pulses due to electron multiplication by the gas are easily identified because they are very different in shape from the direct scintillation light in the liquid.

Typical luminescence signals (S1) are displayed in Figure 2 as a function of time, for a $\beta$ or $\gamma$ track (Figure 2a) and for an Argon ion recoil from a fast neutron (Figure 2b). As discussed later on, the shapes of the scintillation light for the two kinds of events are remarkably different.

The time between the primary pulse (S1) and the secondary pulse (S2) is related to the drift time $T_{drift}$ of the electrons travelling across the liquid (z-coordinate). Given the dimensions of the chamber and the applied electric fields, $T_{drift} < 40$ µs.

The pulse shapes of the 7 photo-multipliers are also individually recorded. Since the secondary pulse (S2) is very near the photo-multipliers, the centroid of the (S2) light, detected by individual PMT's, permits to localise the transverse x-y coordinates of point-like interactions, while, as already pointed out, the z-coordinate is measured by the time difference $T_{drift}$ between the onset of the (S2) and (S1) signals. This technology will be vastly more efficient for the larger size detector under construction where an accuracy of the order of 1 cm is expected. The (S1) and (S2) signals will now be discussed in more detail.

*The (S1) signal.* The main luminescence signal from high density noble liquid scintillation excited by charged particles is in the vacuum ultraviolet region. For Argon it is a broad structure-less band with a width of about 10 nm centred around 128 nm and it has the same spectrum as the so-called second continuum spectrum in gas discharge. Excitation exhibits no significant difference from VUV spectra taken in the high pressure gaseous phase. The origin of the luminescence has been attributed to low excited molecular states (self-trapped excitons), namely to the transitions from the two lowest molecular states ($^1\Sigma_u^+$ and $^3\Sigma_u^+$) to the ground state. The primary scintillation light in liquid Argon is consequently characterised by the presence of two exponential shapes with very different time constants [29]: a fast $^1\Sigma_u^+$ singlet component with $\tau_{singlet}$ = 7.0 ± 1.0 ns, and a slow $^3\Sigma_u^+$ triplet component with $\tau_{triplet}$ = 1.6 ± 0.1 µs[2].

Results show that the decay times for the $^1\Sigma_u^+$ and $^3\Sigma_u^+$ states do not depend on the density of excited species along the particle track [29], i.e. linear energy transfer (LET), although the intensity ratios of the singlet states to the triplet states remarkably depend on LET. Therefore quenching collisions must occur at a very early stage after excitation since the decay times of molecular states would otherwise become shorter under Argon recoil excitation than under $\beta$ or $\alpha$ excitation.

---

[2] We observe that the actual triplet decay time is reduced by relatively small contaminations (few ppm) of N2. As consequence also the ratio of electron and recoil responses is modified. For details, see S. Himi et al. [42] and our own measurements [43].

The striking difference in the pulse shape for events associated to electrons and to ion recoils is visible in Figures 2a and 2b for the pulse (S1).

Let the time integrated number of photoelectrons of each the two components be $I_{singlet}$ and $I_{triplet}$ for the fast singlet state ($^1\Sigma_u^+$) and the slow triplet state ($^3\Sigma_u^+$) respectively. The intensity ratio $I_{singlet}/I_{triplet}$ are found [29] to be 0.3, 1.3 and 3 for electron, $\alpha$-particle and ion recoil excitation, respectively. This result shows the strong enhancement of the $^1\Sigma_u^+$ formation for higher deposited energy densities. In addition to the two above time constants, an intermediate component which has a decay time of 20 ÷ 40 ns is observed [29], with an intensity of the order of 10 ÷ 20 % of the total. This component [26] is not better known at present.

We introduce the pulse shape fast discrimination parameter $F = I_{singlet}/(I_{singlet} + I_{triplet})$. The actual distribution of $F$ follows generally a binomial distribution, representing the statistical fluctuations due to a limited number of photoelectrons. The pulse shape separation is therefore strongly dependent on the number of collected photoelectrons. In order to achieve a meaningful separation the number of collected photoelectrons must be adequate. In this experiment the minimum number of photoelectrons $I_{singlet} + I_{triplet}$ in the (S1) signal is in excess of 50, corresponding to an energy loss for the Argon recoils of 40 keV.

To evidence the necessity of a sufficient light yield in order to perfect the pulse shape separation we have artificially blanked the signals of 4 out of the 7 photo-multipliers. With only 3 photo-multipliers left, a very substantial worsening of the distributions of parameter $F$ is observed, as expected.

In practice, since signals of Figure 2 are already integrated with a time constant of 40 $\mu$s, we extract $F$ from the two time windows after the onset of the (S1) pulse, namely the fast component $A_1$ from the (integrated) pulse after 200 ns and $A_2$ from a second slow/fast component after 5.25 $\mu$s. The actual fast parameter $F$ is then numerically computed as $F = 0.99(A_1/A_2) - 0.118$. The parameter $F$ is found centred on $F = 0.31$ for electrons and on $F = 0.75$ for Argon from neutron recoils (namely $I_{singlet}/I_{triplet} = 3/1$), both with nearly Gaussian narrow distributions.

*The (S2) signal.* As already explained, emitted electrons, which have survived columnar recombination, after drifting across the liquid Argon and traversed the liquid/gas boundary, are multiplied in the vicinity of the grids and emit some additional VUV light delayed by the electron drift time, the so called (S2) signal. Depending on the values of the electric fields, the ratio $(S2)/(S1)$ is typically of the order of 180 for minimum ionizing particles, 3 for $\alpha$ particles, and 10 for Ar-recoils. The phenomenology for the ratio $(S2)/(S1)$ from different origins is rather elaborate and it will be described in detail further on.

To conclude, two independent pulse shape selections are employed simultaneously, namely (1) the shape in time of the primary signal (S1) and (2) the comparison of the primary to the secondary signal coming from the gas, namely $(S2)/(S1)$. As already pointed out, they are both vastly different for $\beta$ or $\gamma$ and for heavily ionising particles ($\alpha$ and Argon nuclear recoils) and their simultaneous signatures provide an over-all very efficient and powerful pulse shape discrimination.

### 3.3. Intensity of scintillation luminescence of nuclear recoils.

Few studies are available in the literature for the luminescence of liquid Argon with particles heavier than $\alpha$-particles. From our data we conclude that the decay times observed from liquid Argon under heavy ion recoil excitation in the energy interval from 20 to 100 $keV_{ion}$ are the same as those measured for instance under $\alpha$-particle excitation. Therefore the states $^1\Sigma_u^+$ and $^3\Sigma_u^+$ (self trapped excitons) are responsible of the luminescence also of heavy ion recoil excitation.

The scintillation (S1) signal produced by ion recoils of a given kinetic energy is expected generally to be smaller than the one due to minimum ionising tracks. In order to determine experimentally the number of photoelectrons (S1) as a function of the kinetic energy $keV_{ion}$ for Argon ions, two measurements have been performed: i) neutrons from an Am-Be source; ii) neutrons from the rock without the heavy shield.

*Neutron calibration with Am-Be source.* Signals from Argon recoils were studied with a dedicated calibration with a weak Am-Be neutron source inserted near the inner detector, inside the shield. Neutron elastic scatterings are an excellent calibration for the study of the response of the detector to low energy Argon recoils. Data were recorded with a drift field 1 kV/cm.

The number of events as a function of the Pulse Shape Discrimination Parameter $F$ is shown in Figure 3 in the recoil energy ranges from 40 keV to 60 keV (Figure 3a) and from 60 keV to 130 keV (Figure 3b). Both all events and those with $10 < (S2)/(S1) < 30$ and $8 < (S2)/(S1) < 22$ are shown respectively for (a) and (b), corresponding to the window of neutron-induced argon recoils. The $(S2)/(S1)$ cut, while strongly depleting the population of electron-like events, leaves the population of Argon ion recoils essentially unaffected.

The observed (S1) distribution for Argon recoils is shown in Figure 4, compared with the results of the Monte Carlo simulation. The observed spectrum shape is well reproduced with a (constant) light yield for Argon recoils from neutrons which is $Y_{Ar}$ = 1.26±0.15 phe/keV. The error is mainly determined by uncertainties in the evaluation of the Monte Carlo simulated neutron flux induced in the chamber, for the computation of which one has to take into account the source intensity and the geometry of propagation of neutrons in the surrounding materials.

*Underground neutrons without the shield.* During the early running, when the large shield around the detector was not yet installed, one could observe a significant rate of neutron events generated by the rocks nearby, amounting to about 8 ev/day above 30 keV in the 2.3 litre chamber. In spite of the significantly smaller statistics, the (S1) distribution is in good agreement with the Monte Carlo simulation of neutron flux from the rock [41] for $Y_{Ar} \approx 1.55 \pm 0.4$ phe/keV.

### 3.4. Ionisation driven signal $(S2)/(S1)$.

The signals from neutron induced Ar-recoils exhibit a rather non trivial pulse shape dependence of the $(S2)/(S1)$ ratio of the extracted electrons to the primary scintillation as a function of the kinetic energy of the recoils.

In the case of electrons of $\approx$ 100 keV, working with the electric field configuration described before, the average ratio is $(S2)/(S1) \approx 150$. The ratio $(S2)/(S1)$ is only very slowly changing with the energy of the electron-like events in the interval of interest, indicating a very good proportionality between the scintillation luminescence and the direct emission of the electrons.

In the case of heavily ionising tracks (i.e. a large LET), a strong columnar recombination takes place and the number of liberated electrons is correspondingly reduced. For instance in the case of $\alpha$-particles from $^{222}$Rn, the ratio $(S2)/(S1)$ is about 50 times smaller than that for minimum ionising particles, providing a very efficient identification of the energy density of excited species along the particle track, i.e. of the linear energy transfer (LET).

Naïve considerations would suggest that in the case of Argon recoils from the liquid, the ratio $(S2)/(S1)$ might be even smaller than with $\alpha$-particles from $^{222}$Rn because of the even higher LET. On the contrary, the experimental observation of neutron induced Argon recoils (25 ÷ 200 keV) shows that: (1) the $(S2)/(S1)$ ratio is much larger than in the case of $\alpha$-particles and (2) it is inversely proportional to the recoil energy, for instance going from $\langle (S2)/(S1) \rangle \sim 30$ at 25 keV to $\langle (S2)/(S1) \rangle \sim 7$ at 150 keV, as shown in Figure 5a and 5b. The separation of these events from the electron signal remains however very large since, as already pointed out, it is $(S2)/(S1) \approx 150$ for electron-like events.

In Figure 5b, the ratio $(S2)/(S1)$ is plotted as a function of the inverse of the recoil energy. For very high energies, the ratio tends to the energy independent limiting case $\langle (S2)/(S1) \rangle \approx 2.1$, which agrees very well with the $^{222}$Rn produced $\alpha$-decays in the volume. But for very low energies $(S2)/(S1)$ is roughly proportional to 1/E, implying the presence of an additional and almost constant electron emission signal in these events, i.e. (almost) irrespective of the actual ion recoil energy (S1).

Neutron induced recoil events are therefore described in a first approximation by a function of the form: $(S2/S1) = a + b/E_{keV}$ where $E_{keV}$ is the (S1) (ion) recoil energy in keV, $a = 2.1$ and $b = 670\ keV$. This function is only slightly depressed for very low recoil energies, near our detection threshold. The actual boundaries corresponding to 90% selection of single neutron induced recoil events[3] are shown in Figures 5a and 5b. The vertical scale in Figure 5b is also expressed in units of the $(S2)/(S1)$ for 100 keV electrons. A very powerful identification is therefore possible in all cases of interest.

Double neutron elastic scatterings may occur within the chamber. When the interaction points are not too close along the z-axis (>1 cm), two distinct secondary pulses are resolved. For this class of "double hit" events (~15% of the cases), the (S1) scintillation luminescence signal corresponds to the sum of the energy depositions. In Figure 5 these events are displayed as the sum of the two (S2) signals vs. the total recoil energy. These events are well separated from the "single hits" and have roughly twice the value of $(S2)/(S1)$ for a given recoil energy, which implies that the constant electron emission signal $b$ is approximately doubled.

Inelastic neutrons, namely channels of the type $n + Ar^{40} \rightarrow n' + Ar^{40} + \gamma's$ are generally open. However the simultaneous presence of $\gamma$ rays introduces variations of both the fast parameter $F$ and of the value of $(S2)/(S1)$, separating experimentally these events from the ones due to purely elastic neutron recoil events.

We shall make the assumption that the perspective weakly interacting WIMP events will follow the same pattern as the one observed by the strongly interacting neutron induced recoils. Therefore the efficiency of the selection cuts for WIMP candidate events is assumed to be equal to that from the neutron calibration data (for energies above 50 keV it is of the order of 90%).

---

[3] The 1/E dependence is well described above about 30 keV; at very low energies a small deviation is apparent; the functional dependence can then be empirically described more precisely as $(S2)/(S1)[E_{keV}] = (a + b/E_{keV})[1 - \exp(E_{keV}/10)]$

## 4.— Results.

We report early data collected with 2.8 x $10^7$ triggers, in a total live time of 52.76 days, corresponding to a total exposure of 96.5 kg x day. Elastic scattering on Argon due to WIMP interactions are selected as single-hits in the relevant energy domain.

The sensitive volume of the chamber is 1.87 litres, corresponding to 2.6 kg. We consider only those events with drift time 10 µs < $T_{drift}$ < 35 µs (fiducial volume cut). This cut excludes: (1) events occurring near the surface of the liquid ($\leq$ 2 cm) where the primary pulse is not sufficiently separated from the secondary for an appropriate event reconstruction; (2) events originating from the contaminations near the cathode region ($\leq$ 0.7 cm).

Taking into account the conical shape of the chamber walls (Figure 1) the resulting fiducial volume is 1.32 litres corresponding to 1.83 kg. The pulses surviving cuts with $(S2)/(S1)$ < 30 and primary pulse fast component $F$ > 0.60 have been visually scanned. This allows to check the overall pulse shape of the primary and secondary pulses and to reject mis-reconstructed or noise events. The resulting distribution of the Log $(S2/S1)$ vs. the Pulse discrimination parameter $F$ is shown in Figures 6 and 7 for the WIMP exposure of 96.5 kg x day (b) and compared with the corresponding distribution of neutron induced ion recoils (a) both for events in the energy interval 40-60 $keV_{ion}$ (Figure 6) and 60-130 $keV_{ion}$ (Figure 7). In the region 40-60 $keV_{ion}$ we observe 8 events of which only 5 stay in the "1-hit" selection band of Figure 5. The largest recorded energy is 54 keV: in the region 60-100 $keV_{ion}$ no event survives. The origin of the events below 55 $keV_{ion}$ is not understood at this stage: they may be of spurious origin, either residual neutrons or e-like background events due to insufficient sensitivity of the rejection criteria. Following the prescription of Ref. [11] the 90% C.L. upper limit has been derived from the observation of zero events in the energy region above 55 $keV_{ion}$. It is shown in Figure 8 where the estimated WIMP-nucleon spin-independent cross section in $cm^2$ vs. an hypothetical WIMP mass in $GeV/c^2$ is also compared with the results coming from the mentioned cryogenic detectors of CDMS [12, 13], EDELWEISS [14,15] and CRESST [16]. In the WIMP scenario we confirm the disagreement with the positive result of DAMA [17], at least as long as a significant contribution is due to the spin independent cross section [20].

A slightly lower limit is also presented in Figure 8 for the case of 40 keV threshold, under the likely assumption that the observed events are due to background. The energy resolution has been taken into account considering the statistical fluctuations in the number of collected photoelectrons and the geometrical effects due to non uniform light collection (~5%), as deduced from calibrations.

The dominant systematic effect is given by the uncertainty on the nuclear recoil light yield. Considering an error of 15% on the light yield, the variation on the limit is of the order of 20% for 100 GeV/$c^2$ WIMP mass, reaching 30% for 50 GeV/$c^2$.

## 5.— Conclusions

These results have been obtained with a provisional filling of the detector with good grade but commercial atmosphere derived Argon, without any "ad hoc" additional purification. The resulting trigger rate is of the order of 6 counts/s, due to several radio-nuclides that are present either in the liquid or eventually in the surrounding walls of the detector, for which no specific attention to material radio-purity has been exercised. In this preliminary phase of development of our novel technology, the presence of such impurities has been indeed beneficial, since it permits to develop the very high rejection capability against spurious events due to backgrounds from $\gamma$ and $\beta$ radiation with a relatively small detector and in a relatively short data taking time.

The main achievements of this small scale prototype detector have been (1) perfecting the sensitivity of LAr ensuring a sufficient luminescence (1.26 phe/keV) for Argon recoils in the energies of interest and (2) to demonstrate the validity of the techniques that will be used in the future. These methods acquire a strong statistical significance provided the number of primary photoelectrons is sufficiently high (typically > 50 phe for kinetic Argon recoil energies > 40 keV). In particular the simultaneous introduction in actual underground conditions of the two independent discrimination methods, namely the pulse shape discrimination parameter $F$ and the ratio $(S2)/(S1)$ between the delayed electrons signal extracted from the liquid to the gas (S2) and the initial luminescence photoelectron yield (S1), appear to be absolutely necessary in order to ensure a sufficient robustness to the identification of a possible WIMP signal. This primes the choice of Argon [36] with respect to our earlier studies performed with Xenon.

These are only the first results and the data taking is continuing. Notwithstanding the small 2.3 litre table-top test chamber has already permitted to reach a sensitivity limit against WIMP which is quite comparable with the one of traditional cryogenic detectors [12-16] at mK temperatures.

In order to further improve the sensitivity of the method, besides increasing the active mass of a new detector, we intend to achieve:

- A reduction of the rate of background counts due to $\beta$ and $\gamma$ both in the liquid and in the surrounding structures.

- A much stronger discrimination against the background with the addition of an active anticoincidence volume rather than only of a passive shielding.

- Rejection of most of the backgrounds occurring near the walls of the sensitive volume by means of the x-y components of the signals (in addition to the z-component already determined by time of flight) provided by the much larger number of photo-multipliers, capable to locate the centroid of the S2 signal with an accuracy of about 1 cm.

**Further information added in proof.**

After the completion of this report, data taking has been resumed in the period September-December 2006 with significant results. In particular:

The electronics for the pulse height distributions has been incremented from 20 MHz to 100 MHz and the electronics of the front end has been improved. An additional sample of 43 kg x day has been collected. The combined rejection power for electrons and gamma's due to the combined effects of fast parameter $F$ and of the value of $(S2)/(S1)$ has been significantly improved.

With an acceptance of 50% for nuclear recoils, the measured discrimination power for events above 35 photoelectrons is now better than $3 \times 10^{-7}$. When raising the threshold to 50 photoelectrons, the measured discrimination for a 70% nuclear recoils acceptance is $3 \times 10^{-7}$ and the extrapolated discrimination for a 50% acceptance is better than $1.0 \times 10^{-8}$.

The new results hint at the fact that the two discrimination methods are independent within the observed statistics. All events and the events with $1.0 < \text{Log}(S2/S1) < 1.5$ are indeed quite similar and show no sign of correlation.

One preminent nuclear recoil candidate has been recorded inside the red box defined by a ~70% acceptance for nuclear recoils, with its position on the very bottom of the 70% acceptance region. The origin of such event is under investigation.

An isotopically separated Argon filling (with an $^{39}$Ar residual contamination better than 2% of the natural content) has been successfully produced [39] and it will be operational soon in the 2.3 liter chamber. This will strongly reduce the electron and gamma backgrounds produced in the liquid, presently dominated by the presence of $^{39}$Ar.

## 6.— Acknowledgements.

We wish to thank INFN for the continuous support of this initiative since the very beginning of this development. We thank in particular the INFN President, Prof. Roberto Petronzio, for his specific attention and for his personal involvement in pursuing and sustaining our efforts and the LNGS Director, Prof. Eugenio Coccia for his continuous support and encouragement. We acknowledge Prof. Enrico Bellotti, Dr. Oliviero Cremonesi and Prof. Flavio Gatti who have acted as referees of INFN-Commissione II.

This experiment is a direct spin-off of major technological advances brought about by the ICARUS collaborators, without whom this work would have been impossible.

A.M. Szelc has been in part supported by a grant of the President of the Polish Academy of Sciences and by the MNiSW grant 1P03B04130.

The WARP programme is supported since March 2006 by the National Science Foundation under Grant no. PHY-0603376. Funding from the NSF enabled the construction of the upgraded electronics, which permitted to obtain the latest results on pulse shape discrimination.

We acknowledge the dedication and professionalism of the personnel of the INFN mechanical and electronics workshops that have designed and built the components of the 2.3 litre detector, in particular Angelo Freddi, Claudio Scagliotti and Filippo Vercellati from the Pavia mechanical workshop; Orlando Barnabà, Giulio Musitelli and Roberto Nardò from the Pavia electronic workshop.

We wish to acknowledge the many people from LNGS staff that helped us with the installation and run of the chamber; especially Ing. Piergiorgio Aprili, Ing. Alfredo Fulgenzi, Ing. Paolo Martella, Ing. Roberto Tartaglia for the logistic of underground operation; the LNGS chemical service and in particular Ing. Marco Balata and Luca Ioannucci, for support in the cryogenics; the LNGS mechanical and electronic services, and in particular Ercolino Tatananni and Marco D'Incecco; Bert Harrop from the Princeton group for electronics. A special thank goes to Paolo Cennini for the many years of highly qualified and very effective support both to ICARUS and to WARP.

We warmly thank Dr. Alfredo Ferrari for his help in several FLUKA simulations of our detector; from the Princeton group, Yue Zhao, John Appel, David Krohn, Ben Loer, Daniel Marks and Richard Saldanha for analysis.

We finally thank Dr. Marco Roncadelli for the many and very interesting discussions about the theory and phenomenology of Dark Matter.

## 7.— References.

## 8.— Figure captions.

Figure 1.  Layout of the test chamber. 1) Liquid Argon Drift volume; 2) Reflector/Wavelength shifter layer; 3) Photo-multiplier; 4) Heating resistance; 5) HV supply; 6) Vacuum port; 7) LAr in; 8) Filter (for filling); 9) Filter (for recirculation); 10) External Argon dewar.

Figure 2.  Two typical time recordings of the integrated scintillation signal (S1) from liquid Argon excited by charged particles. In Figure 2a it is shown a typical electron-like signal and in Figure 2b a neutron induced Argon recoil. The experimentally observed signal has been decomposed into the two components $I_{singlet}$ and $I_{triplet}$ due to the fast singlet state ($^1\Sigma_u^+$) and the slow triplet state ($^3\Sigma_u^+$) respectively. The sums are also shown, in excellent agreement with the observed signals.

Figure 3.  Distribution of the Pulse Shape Discrimination Parameter ($F$) with an AmBe neutron source in the recoil energy range from 40 keV to 60 keV (a) and from 60 keV to 130 keV (b). The blue dots represent all events, while the red points refer to a population having $10 < (S2)/(S1) < 30$ and $8 < (S2)/(S1) < 22$ respectively for (a) and (b), corresponding to the window appropriate for neutron-induced argon recoils. The (S2)/(S1) cut, while strongly depleting the population of electron-like events, leaves the population of Argon ion recoils essentially unaffected.

Figure 4.  Energy spectrum of the Recoil-like events in the fiducial volume collected in the calibration with a weak AmBe neutron source. The data (red dots) are compared with the Monte Carlo prediction. The energy fit gives a conversion factor of 1.26 ± 0.15 photoelectrons/keV.

Figure 5.  Distribution of the Argon recoil signals from the Am-Be source calibration in the plane (S2)/(S1) vs. energy E (a), and vs. 1/E (b). The 1-hit selection band defined in the text is drawn. Red dots are the events contained in the single-hit selection band (90%). The lowest threshold used for the analysis is shown.

Figure 6.  (a) Distribution in the plane (S2)/(S1) vs. Pulse Shape Discrimination Parameter of the events obtained with an Am-Be source calibration of the WARP detector in the recoil energy range 40-60 keV; (b) distribution in the same energy range obtained from the WIMP exposure of 96.5 kg day. The box shown is indicative: out of the 8 events which are observed, only 5 belong to the "1-hit" selection band of Figure 5.

Figure 7.  Same distributions as in Figure 6 but in the energy interval 60-130 keV.

Figure 8.  90 % C.L. spin independent limits (solid blue curve) obtained by WARP-2.3l for a total fiducial exposure of 96.5 kg·d and a threshold of 55 keV. An analogous curve for $E_{rec} > 40$ keV under the (optimistic)

assumption that the observed 5 events are due to background is also plotted (dashed blue curve). The limit is compared with the ones from CDMS [12, 13], EDELWEISS [14,15] and CRESST [16] and with the allowed $3\sigma$ C.L. from the DAMA 1-4 annual modulation data [17].

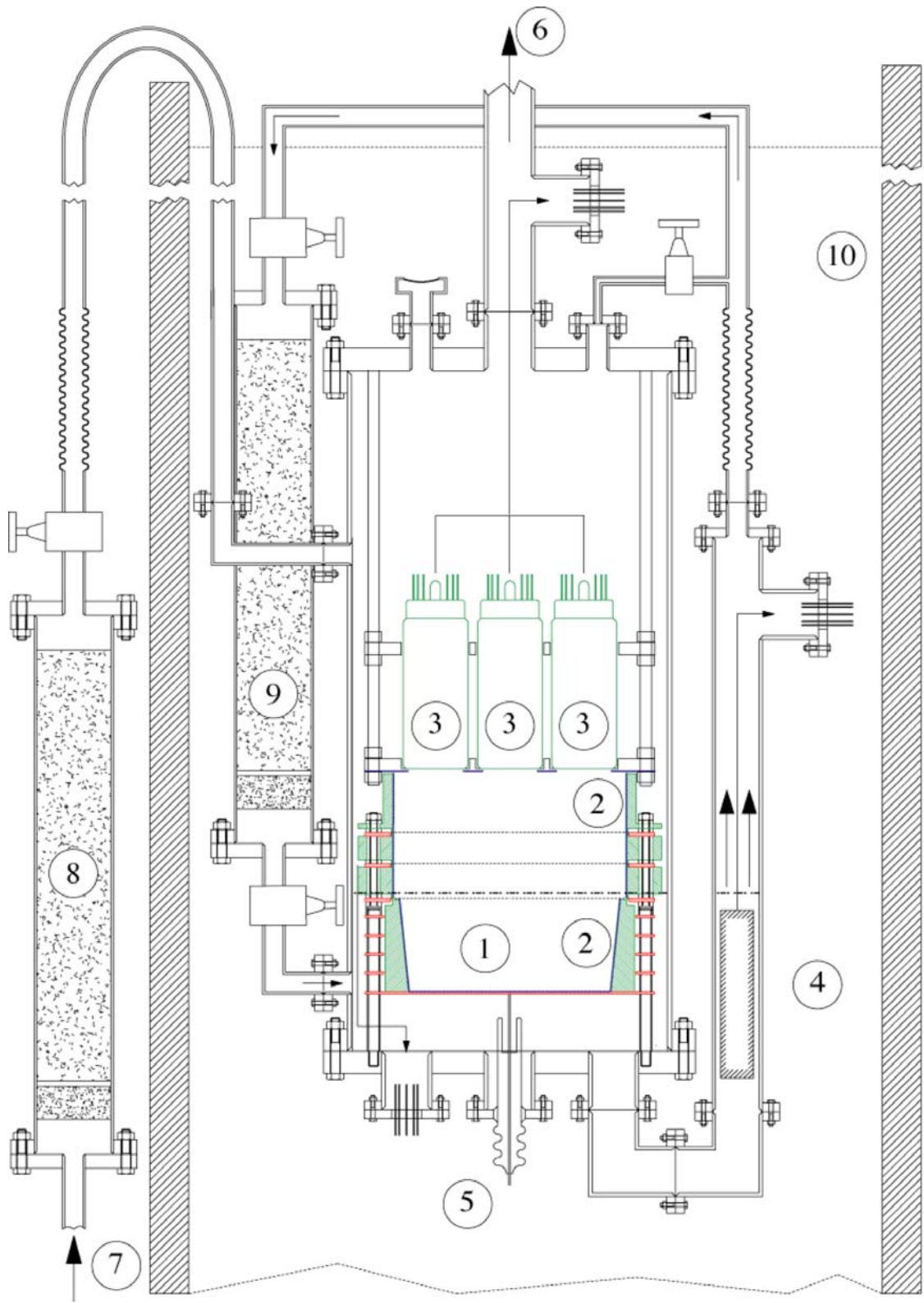

Figure 1.

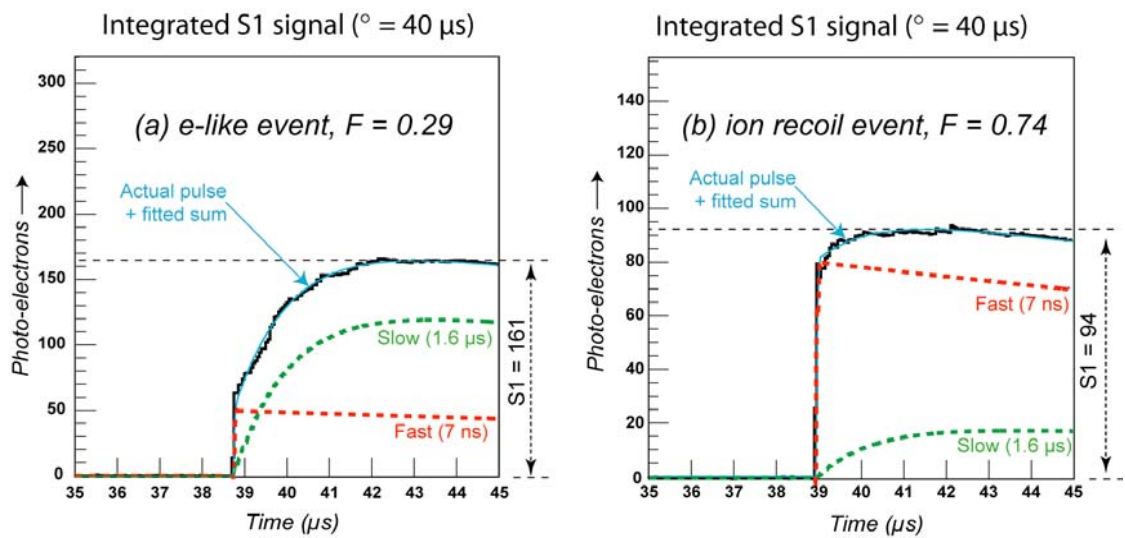

Figure 2.

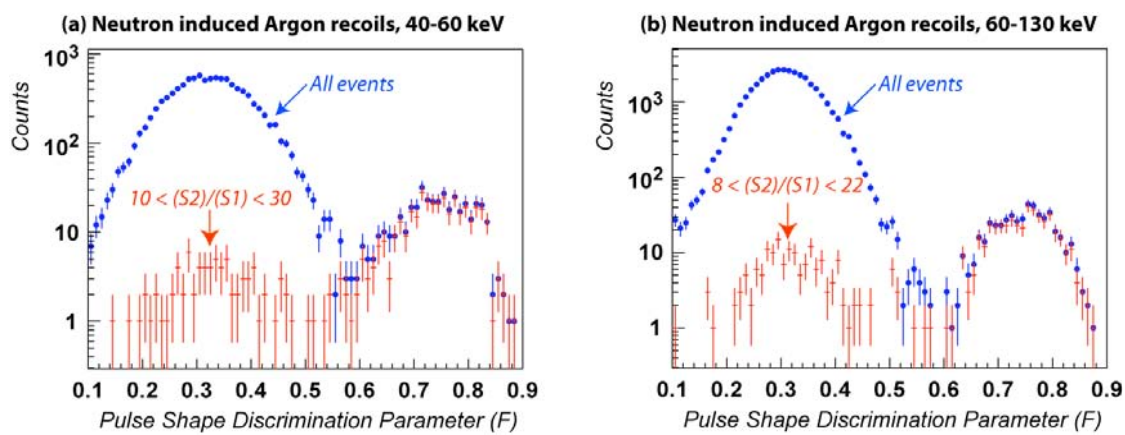

Figure 3.

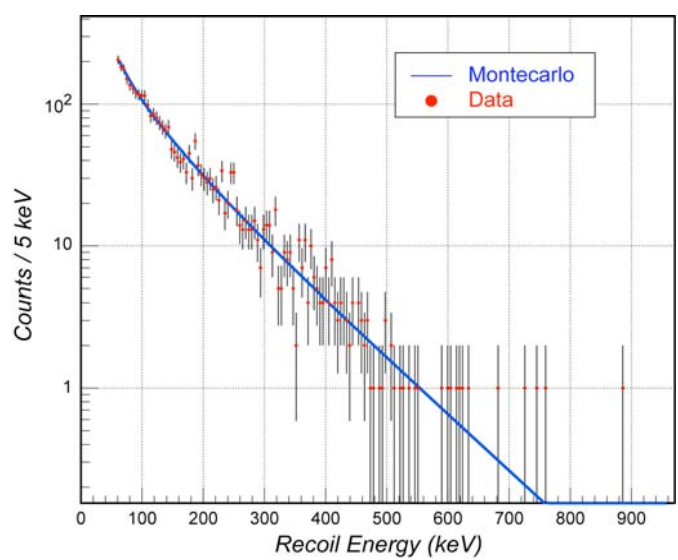

Figure 4.

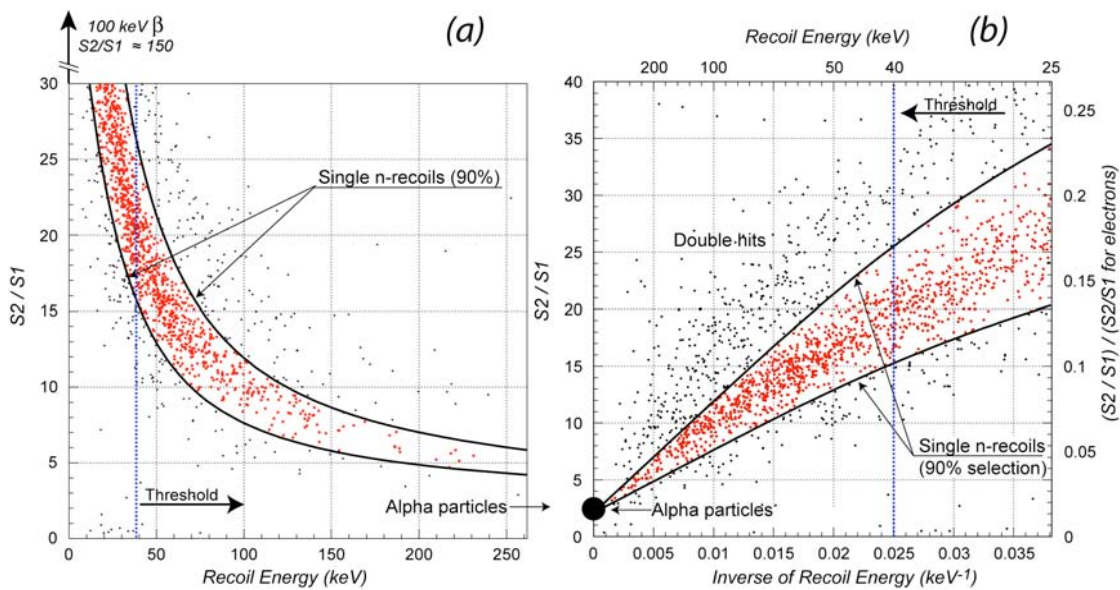

Figure 5

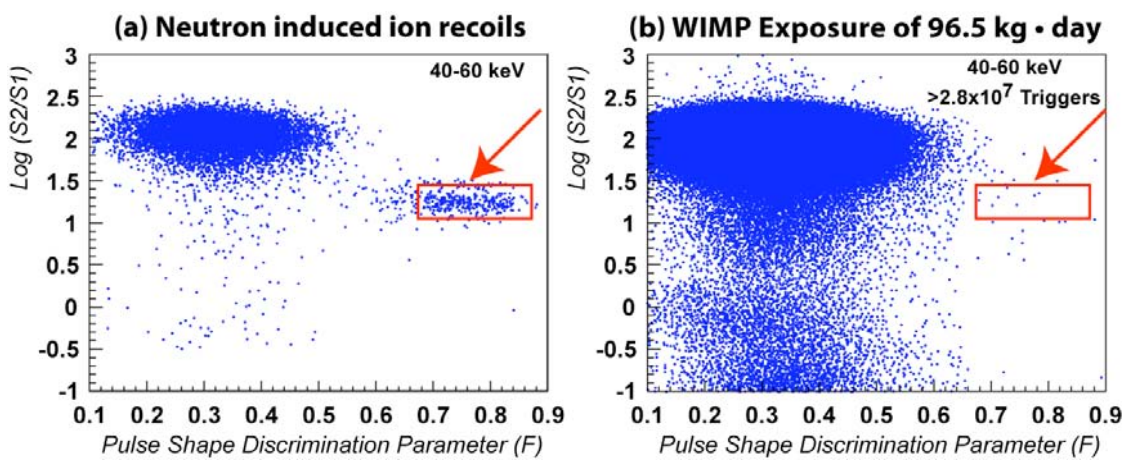

Figure 6

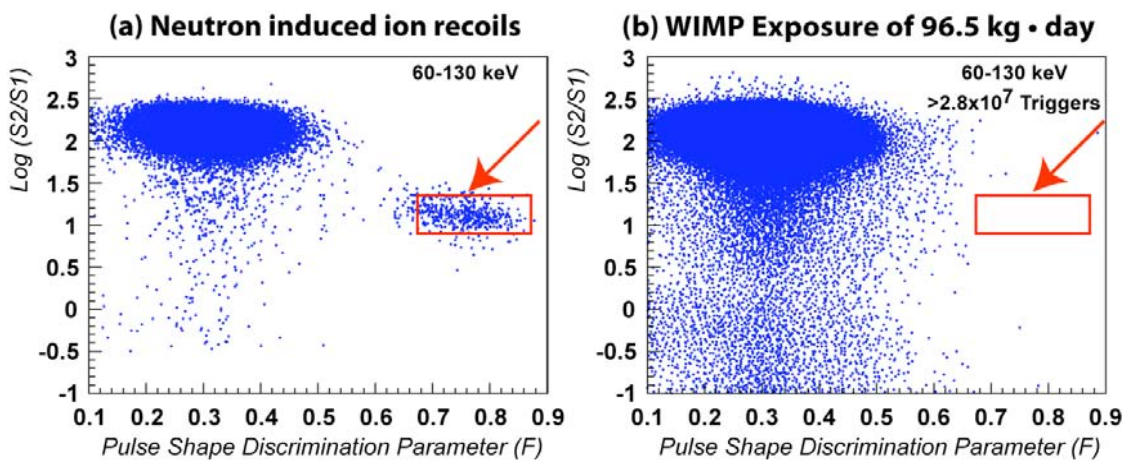

Figure 7

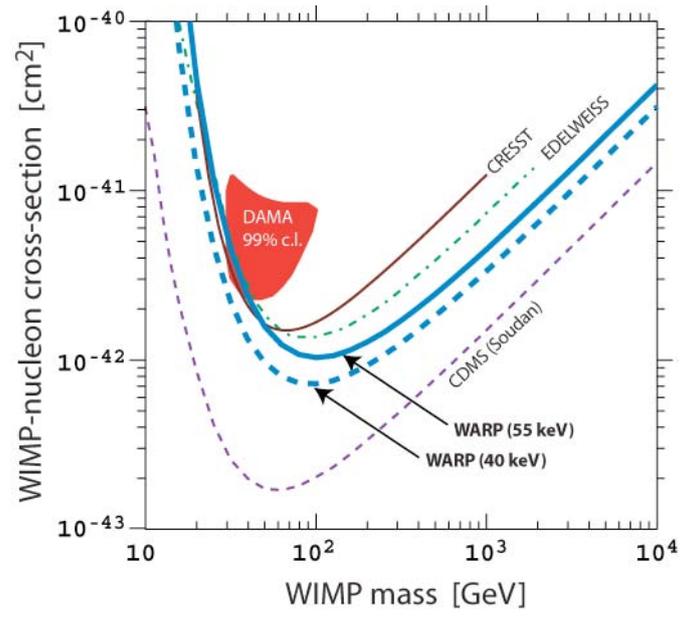

Figure 8